\documentclass[preprint2, times, tighten, twocolappendix]{aastex631}

\usepackage{makecell}
\usepackage{ragged2e}
\usepackage{hyperref}
\usepackage{CJK}

\providecommand{\e}[1]{\ensuremath{\times 10^{#1}}}

\begin{document}

\title{JWST Advanced Deep Extragalactic Survey (JADES) Data Release 5: \\MIRI Coordinated Parallels in GOODS-S and GOODS-N}

\author[0000-0002-0786-7307]{Stacey Alberts}
\affiliation{Steward Observatory, University of Arizona, 933 North Cherry Avenue, Tucson, AZ 85719, USA}
\affiliation{AURA for the European Space Agency (ESA), Space Telescope Science Institute, 3700 San Martin Dr., Baltimore, MD 21218, USA}
\email{salberts@stsci.edu}


\author[0000-0002-2929-3121]{Daniel J.\ Eisenstein}
\affiliation{Center for Astrophysics $|$ Harvard \& Smithsonian, 60 Garden St., Cambridge, MA 02138, USA}

\author[0000-0002-8651-9879]{Andrew J.\ Bunker}
\affiliation{Department of Physics, University of Oxford, Denys Wilkinson Building, Keble Road, Oxford OX1 3RH, UK}

\author[0000-0002-9551-0534]{Emma Curtis-Lake}
\affiliation{Centre for Astrophysics Research, Department of Physics, Astronomy and Mathematics, University of Hertfordshire, Hatfield AL10 9AB, UK}

\author[0009-0009-8105-4564]{Qiao Duan}
\affiliation{Kavli Institute for Cosmology, University of Cambridge, Madingley Road, Cambridge, CB3 0HA, UK
Cavendish Laboratory, University of Cambridge, 19 JJ Thomson Avenue, Cambridge, CB3 0HE, UK}

\author[0000-0003-4565-8239]{Kevin Hainline}
\affiliation{Steward Observatory, University of Arizona, 933 North Cherry Avenue, Tucson, AZ 85719, USA}

\author[0000-0002-8543-761X]{Ryan Hausen}
\affiliation{Department of Physics and Astronomy, The Johns Hopkins University, 3400 N. Charles St., Baltimore, MD 21218}

\author[0000-0003-4337-6211]{Jakob M.\ Helton}
\affiliation{Department of Astronomy \& Astrophysics, The Pennsylvania State University, University Park, PA 16802, USA}

\author[0000-0001-7673-2257]{Zhiyuan Ji}
\affiliation{Steward Observatory, University of Arizona, 933 North Cherry Avenue, Tucson, AZ 85719, USA}

\author[0000-0002-9280-7594]{Benjamin D.\ Johnson}
\affiliation{Center for Astrophysics $|$ Harvard \& Smithsonian, 60 Garden St., Cambridge, MA 02138, USA}

\author[0000-0002-6221-1829]{Jianwei Lyu (\begin{CJK}{UTF8}{gbsn}吕建伟\end{CJK})}
\affiliation{Steward Observatory, University of Arizona, 933 North Cherry Avenue, Tucson, AZ 85719, USA}

\author{Jane Morrison}
\affiliation{Steward Observatory, University of Arizona, 933 North Cherry Avenue, Tucson, AZ 85719, USA}

\author[0000-0003-4528-5639]{Pablo G. P\'erez-Gonz\'alez}
\affiliation{Centro de Astrobiolog\'ia (CAB), CSIC-INTA, Ctra. de Ajalvir km 4, Torrej\'on de Ardoz, E-28850, Madrid, Spain}

\author[0000-0003-2303-6519]{George H. Rieke}
\affiliation{Steward Observatory, University of Arizona, 933 North Cherry Avenue, Tucson, AZ 85719, USA}

\author[0000-0002-7893-6170]{Marcia Rieke}
\affiliation{Steward Observatory, University of Arizona, 933 North Cherry Avenue, Tucson, AZ 85719, USA}

\author[0000-0002-5104-8245]{Pierluigi Rinaldi}
\affiliation{Space Telescope Science Institute, 3700 San Martin Drive, Baltimore, Maryland 21218, USA}

\author[0000-0002-4271-0364]{Brant Robertson}
\affiliation{Department of Astronomy \& Astrophysics, University of California, Santa Cruz, 1156 High Street, Santa Cruz, CA 95064, USA}

\author[0000-0001-6561-9443]{Yang Sun}
\affiliation{Steward Observatory, University of Arizona, 933 North Cherry Avenue, Tucson, AZ 85719, USA}

\author[0000-0002-8224-4505]{Sandro Tacchella}
\affiliation{Kavli Institute for Cosmology, University of Cambridge, Madingley Road, Cambridge, CB3 0HA, UK}
\affiliation{Cavendish Laboratory, University of Cambridge, 19 JJ Thomson Avenue, Cambridge, CB3 0HE, UK}

\author[0000-0003-2919-7495]{Christina C.\ Williams}
\affiliation{NSF National Optical-Infrared Astronomy Research Laboratory, 950 North Cherry Avenue, Tucson, AZ 85719, USA}

\author[0000-0001-9262-9997]{Christopher N.\ A.\ Willmer}
\affiliation{Steward Observatory, University of Arizona, 933 North Cherry Avenue, Tucson, AZ 85719, USA}

\author[0000-0002-8876-5248]{Zihao Wu}
\affiliation{Center for Astrophysics $|$ Harvard \& Smithsonian, 60 Garden St., Cambridge, MA 02138, USA}






\begin{abstract}
Medium to ultra-deep mid-infrared imaging surveys with the James Webb Space Telescope (JWST)'s Mid-Infrared Instrument (MIRI) are reframing our view of the early Universe, from the emergence of ultra-red dusty and quiescent galaxies to the epoch of reionization to the first galaxies.  Here we present the MIRI coordinated parallels component of the JADES program, which obtained ultra-deep ($155$ ks) imaging at $7.7\,\mu$m over $\sim10$ arcmin$^2$ as well as medium depth ($\sim5-15$ ks) imaging at $7.7,\,12.8$, and $15\,\mu$m over $\sim36$, 25, and 22 arcmin$^2$, respectively, in the GOODS-S and GOODS-N fields.  This paper describes the data reduction, which combines the official JWST Calibration Pipeline with custom steps to optimize flagging of warm/hot pixels and optimize background subtraction.  We further introduce a new step to address artifacts caused by persistence from saturating sources.  The final, fully reduced JADES/MIRI mosaics are being released as part of JADES Data Release 5, along with prior-based forced photometry using NIRCam detection images, providing critical rest-frame near-infrared and optical constraints on early galaxy populations.

\end{abstract}



\section{Introduction} \label{sec:intro}

The James Webb Space Telescope \citep[JWST;][]{rigby2023, gardner2023} has dramatically advanced our understanding of galaxy evolution by providing access to the information-rich rest-frame optical and near-infrared wavelengths into the early Universe.  These wavelength regimes encode critical diagnostics for key components of a galaxy's ecosystem: old and intermediate-age stellar populations, recent star formation, metals and dust content, and the presence of active black holes, to name a few. 

\begin{figure*}[ht!]
    \centering
    \includegraphics[width=2.2\columnwidth]{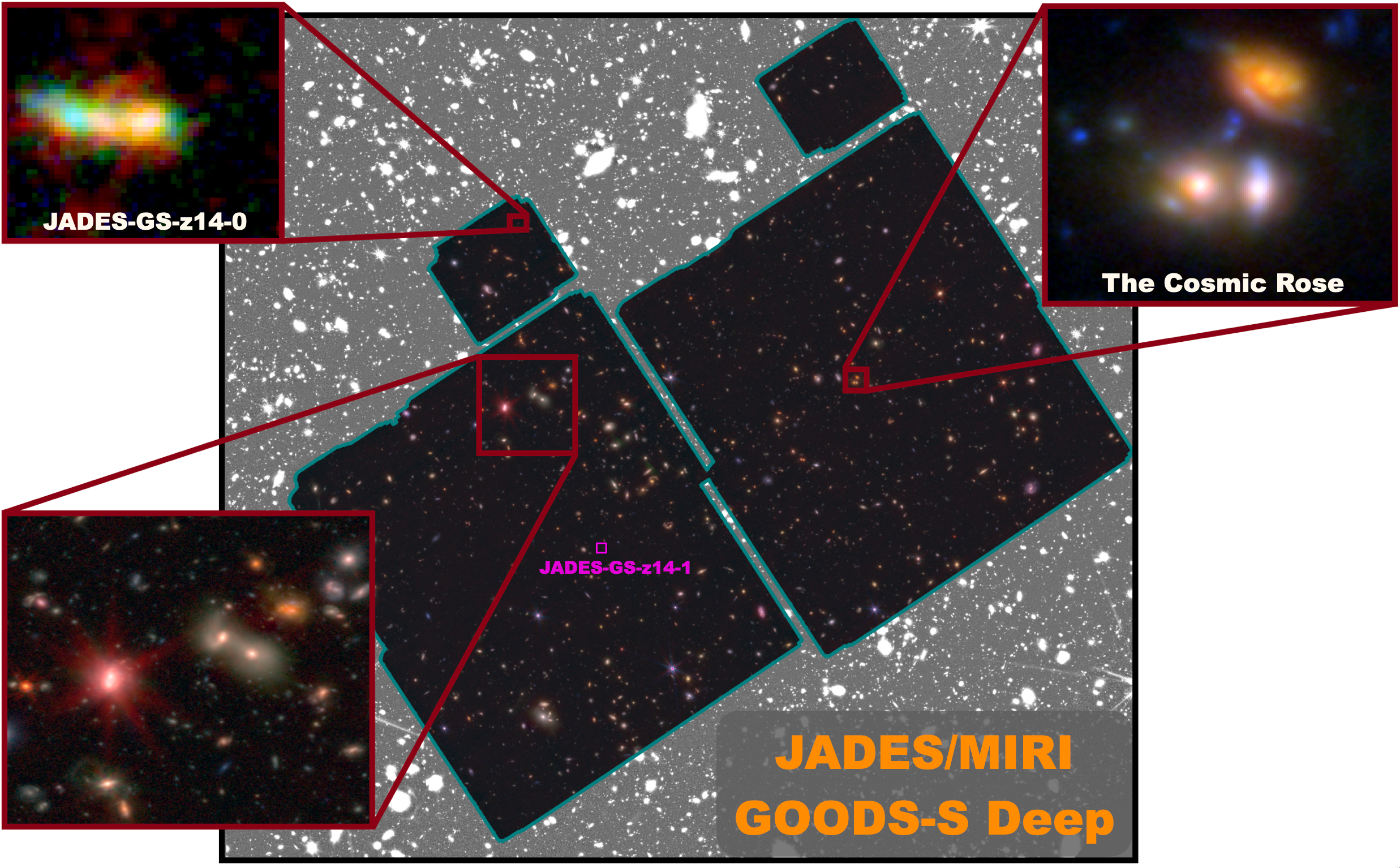}
    \caption{RGB image of the JADES/MIRI GOODS-S Deep parallel (R=MIRI F770W, G=NIRCam F444W, B=NIRCam F200W) shown outlined in teal. The GS-Deep mosaic covers $\sim10$ arcmin$^2$ with an average exposure time of 155 ks (43 hr) in the F770W filter (Section~\ref{sec:deep}). Background grayscale image is a stack of the NIRCam F444W and F200W mosaics. Zoom-in regions: (top left) a $z\sim14$ galaxy detected at F770W \citep[JADES-GS-z14-0;][]{helton2025a, helton2025b}; (top right) a compact grouping of quenched and star forming galaxies at $z\sim3.7$ termed the Cosmic Rose \citep{alberts2024b}; (bottom left) a region containing a representative cross-section of the galaxies in the GS-Deep parallel.  The small purple box highlights JADES-GS-z14-1 \citep[][]{wu2025}, which is undetected in F770W despite falling in a strip of the parallel with a total exposure time of 253 ks (70.7 hr, see Figure~\ref{fig:exp_time}). }
    \label{fig:rgb}
\end{figure*}

The ultra-sensitive imaging and spectroscopy provided by JWST's near-infrared (IR) instruments have lead the charge in building our new view of the early Universe.  However, critical spectral features begin to shift out of the observed-frame near-IR at higher redshifts: the rest-frame near-IR is probed by the observed mid-IR at $z\gtrsim4$ and important optical features such as the H$\alpha$, H$\beta$, and [O{\sc III}] emission lines move into the mid-IR starting at $z\sim6.5-9$.  Following in the footsteps of the Spitzer Space Telescope \citep{werner2004, gehrz2007}, which provided the first rest-frame optical constraints on the stellar populations of galaxies in the Epoch of Reionization \citep[EoR; e.g.,][]{eyles2005, stark2013, duncan2014}, this gap can be filled via deep imaging and spectroscopy with JWST's Mid-Infrared Instrument \citep[MIRI;][]{rieke2015, wright2023}, which provides continuous coverage over $\sim5-29\,\mu$m.

To this end, the first few observing cycles of JWST have included a handful of  medium-depth ($\gtrsim5$ ks) and ultra-deep ($\gtrsim100$ ks) MIRI imaging surveys \citep[see][for a summary through Cycle 3]{alberts2024a}, including the JWST Advanced Deep Extragalactic Survey \citep[JADES;][]{bunker2020, rieke2020, eisenstein2023a}, which is the focus of this paper.  These surveys have primarily observed with the shorter wavelength MIRI filters \citep[e.g.,][]{ostlin2025} to maximize sensitivity and avoid the steeply rising thermal background at $\gtrsim15\,\mu$m \citep{rigby2022a}.  Additionally, due to the small field-of-view (FOV) of the MIRI imager ($\sim2.3$ arcmin$^2$) and relatively large exposure time requirements, deep MIRI surveys to date cover small areas for reasons of practicality; nevertheless, they have yielded significant (and sometimes unexpected) returns.  

Early efforts with MIRI focused on verifying measurements of basic galaxy properties at high redshift by incorporating the rest-frame near-IR and/or optical.  Early analyses found that the inclusion of $\lambda_{\mathrm{obs}}$ $>5\,\mu$m observations with MIRI reduced stellar mass and star formation rate (SFR) estimates by factors of $2-3$ at $z\sim7-9$, in part due to emission line boosting in $\sim4\,\mu$m bands \citep{papovich2023, wang2025}.  Subsequent analyses, however, used deep MIRI imaging to confirm that densely-spaced multi-band NIRCam imaging can mitigate these effects in typical high-redshift galaxies.  For example, \citet{helton2025} examined 47 galaxies at $z\sim8$ using ultra-deep JADES imaging from NIRCam plus MIRI's F770W filter (Figure~\ref{fig:rgb}), finding that $8-14$ NIRCam bands (including some medium bands) can return identical stellar properties and star formation histories with and without the MIRI photometry for $\sim80\%$ of their sample. A similar result was found at $z\sim3-6$ for measurements of basic stellar properties and specifically for commonly used color selections of quiescent and post-starburst galaxies, where MIRI provides the rest-frame J-band anchor typically used to distinguish between older and dusty populations \citep{alberts2024b}.

On the other hand, beyond basic properties of typical galaxies, MIRI has proven to provide unique constraints in specific populations.  MIRI imaging of Little Red Dots \citep[e.g.,][]{labbe2023a} has provided key constraints on the amount of hot dust, ruling out several explanations for their peculiar spectral energy distributions \citep[SEDs; e.g.,][]{williams2023, setton2025}.  In other early galaxies, MIRI has revealed extremely red rest-frame near-IR colors \citep[e.g.,][]{rinaldi2025} and candidates for high-$z$ obscured AGN \citep{lyu2024}. In addition to its wavelength coverage, MIRI's spatial resolution ($\sim0.2-0.85\arcsec$) enables resolved rest-frame near-IR and optical analyses at high redshift \citep{costantin2025}; recently, JADES imaging at F1500W revealed \textit{resolved} hot dust in a Lyman-$\alpha$ continuum leaker at $z\sim3.8$, offset from the Lyman continuum and H$\alpha$ emitting regions \citep{ji2025}.

The most unexpected result from deep MIRI imaging surveys, however, has to be their ground-breaking role in understanding the earliest galaxies.  Despite relatively wide bandpasses, intense emission lines boosting the flux in MIRI imaging filters can provide valuable constraints on optical lines such as H$\alpha$ at $z\gtrsim7$ \citep{rinaldi2023, rinaldi2023a}.  The most striking example of this was presented by \citet{helton2025a}: a record-breaking, luminous $z=14.18$ galaxy \citep[JADES-GS-z14;][]{carniani2024} was detected with the F770W filter at the edge of the ultra-deep JADES MIRI parallel (Figure~\ref{fig:rgb}) due to the $7.7\,\mu$m flux being boosted by [O{\sc III}]+H$\beta$.  This source has now been followed up with MIRI's Low Resolution Spectrograph (LRS), which provides coverage of [O{\sc III}], H$\beta$, and H$\alpha$, three optical lines critical to understanding physics in the first galaxies \citep{helton2025b}.   Serendipitously, a second, fainter $z\sim14$ galaxy \citep[JADES-GS-z14-1;][]{wu2025} was identified in the deepest part of the JADES ultra-deep F770W parallel, where a full 70.7 hr exposure time was used to place a strong upper limit on its $7.7\,\mu$m flux.  This limit constrains the possible equivalent width of [O{\sc III}]+H$\beta$, confirming that this fainter, lower mass $z=13.86$ galaxy lacks the extreme nebular emission that appears common in the luminous galaxies JWST has revealed at $z>10$ \citep[e.g.,][]{bunker2023, deugenio2024, hsiao2024, alvarez-marquez2025, zavala2025, naidu2025}.

In this work, we present the overview, data reduction, and release of deep MIRI imaging obtained as part of JADES.  Totaling roughly 770 hours of observing time allocated to the Near-Infrared Camera \citep[NIRCam;][]{rieke2023} and Near-Infrared Spectrograph \citep[NIRSpec;][]{jakobsen2022, ferruit2022} Guaranteed Time Observation (GTO) teams, JADES dedicated $\sim200$ hours of science time to MIRI coordinated parallels in the GOODS-N and GOODS-S fields.  This time was split into two components: first, an ultra-deep survey across four pointings ($\sim$10 arcmin$^2$) in F770W observed for 155 ks, reaching a a point source sensitivity of 19 nJy ($5\sigma$).  Only one other program, MIDIS \citep{ostlin2025}, has obtained similarly deep MIRI imaging in single pointings of the F560W and F770W bands.  Secondly, JADES/MIRI includes a medium-depth ($5-15$ ks) component covering a non-continguous area of roughly 26 arcmin$^2$ with F770W, F1280W, and/or F1500W.  Comparable medium-depth MIRI imaging surveys include prime MIRI imaging in CEERS \citep{yang2023a} and lensing clusters Abell 2744 and MACS 0416 (PID 5578, PIs E. Iani \& P. Rinaldi), as well as MIRI parallels to the JADES Origins Field \citep{eisenstein2023} and NIRCam/WFSS follow-up of the SMILES field (PID 4549, PI G. Rieke).  As of now, this relatively small number of deep MIRI imaging programs, joined by a growing number of ultra-deep MIRI spectroscopic programs \citep[e.g.,][]{zavala2025, helton2025}, represents our best window into the rest-frame NIR and optical in the EoR and beyond.

This paper is organized as follows: we start with the survey design and data acquisition for the JADES/MIRI parallels in the GOODS-S and GOODS-N fields (Section~\ref{sec:acq}).  This is followed by a description of the data reduction (Section~\ref{sec:data_reduction}), which builds off of the MIRI data reduction pipeline presented for SMILES in \citet{alberts2024a}, here expanded to include a correction for artifacts caused by persistence and modified to address unique aspects of this dataset. Finally, we close with a description of the MIRI mosaics component of JADES Data Release (DR) 5 (Section~\ref{sec:data_release}).  Descriptions of the NIRCam mosaics and photometric catalogs, including NIRCam-based forced photometry in the MIRI bands, slated for release in DR 5 can be found in \citet{Johnson2026} and \citet{Robertson2026}, respectively.  All magnitudes in this work are in AB units \citep{oke1983}.

\begin{figure*}[tbh!]
    \centering
    \hspace{5mm}\includegraphics[width=1.5\columnwidth]{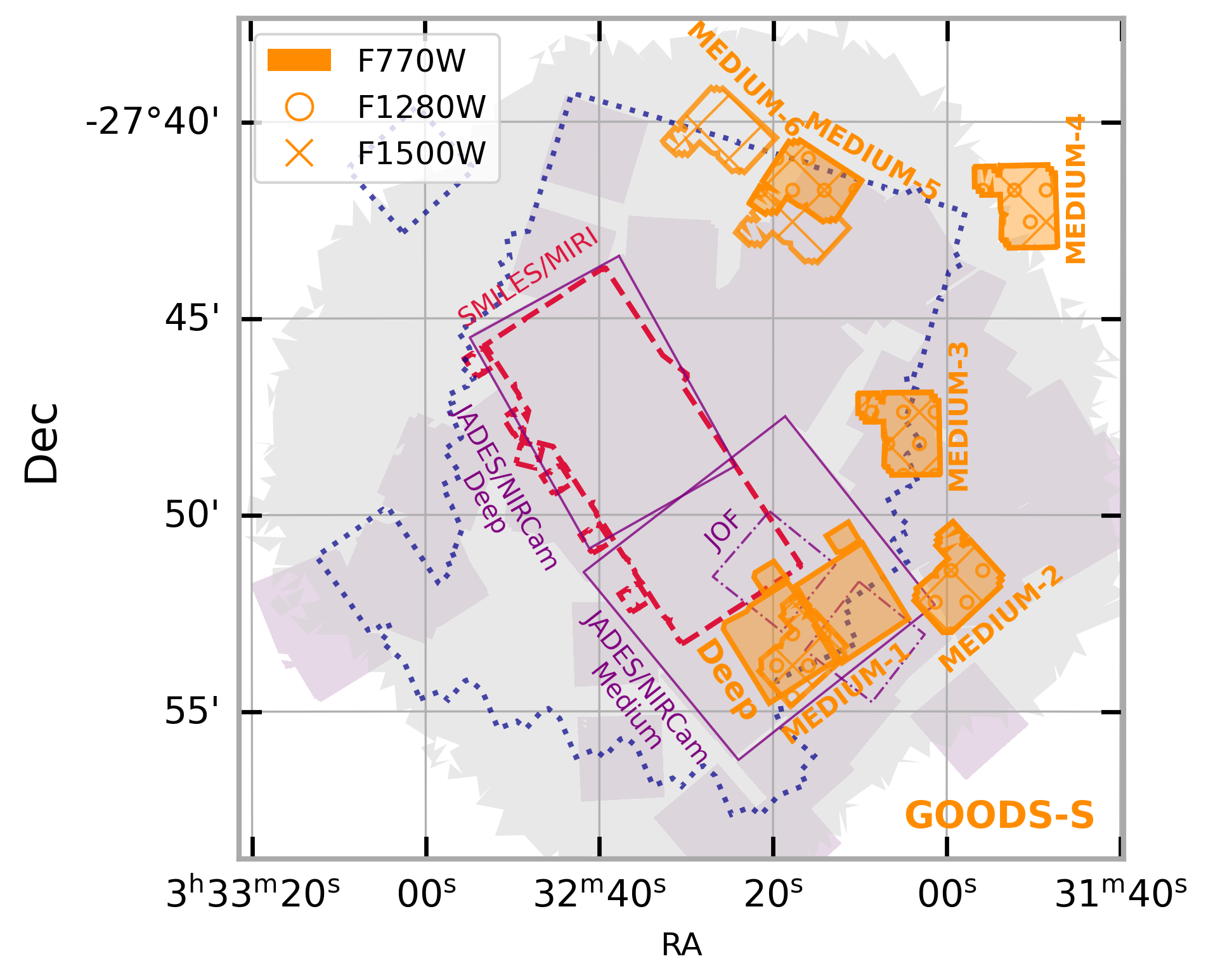}
    \includegraphics[width=1.4\columnwidth]{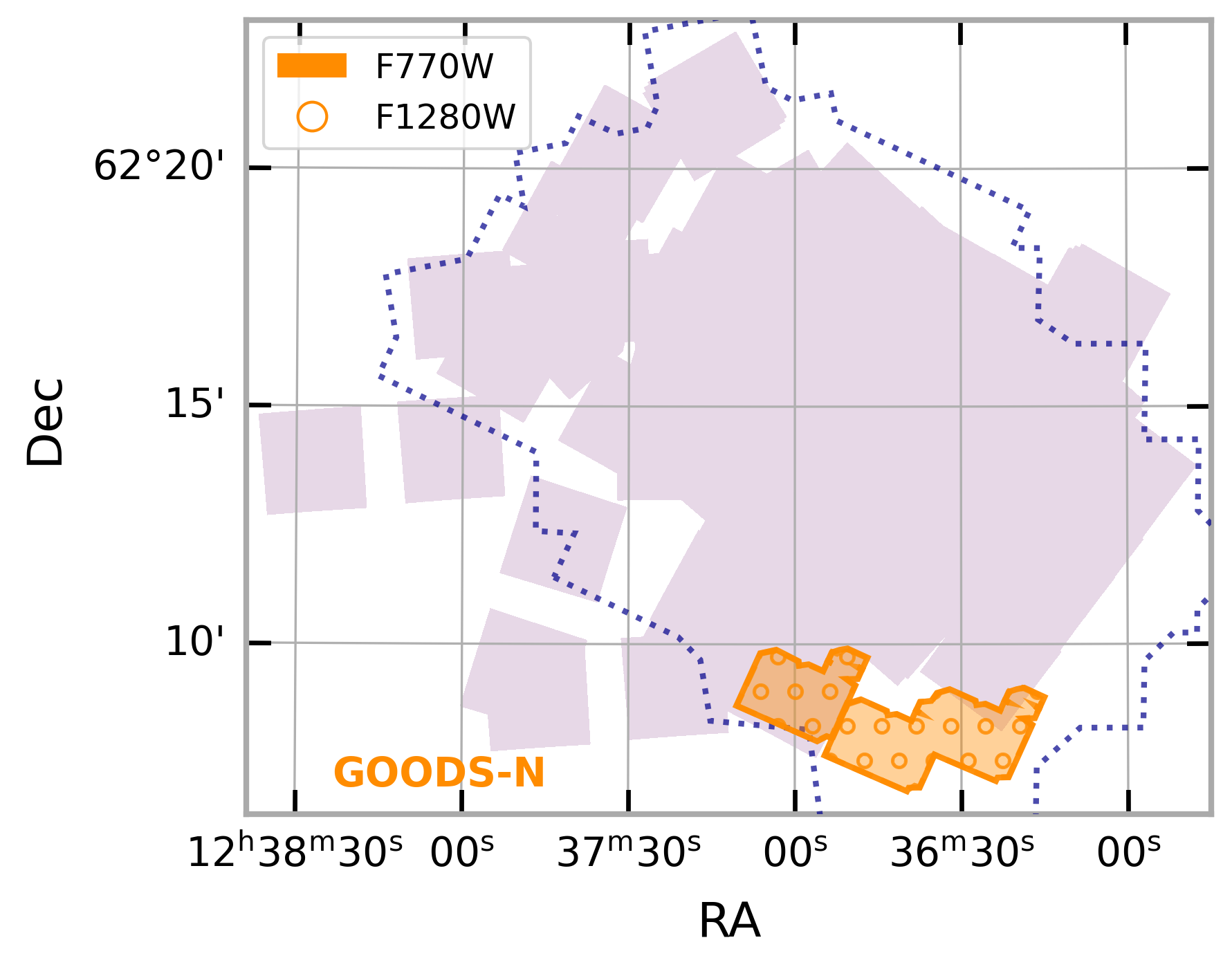}
    \caption{The JADES/MIRI parallel footprints in the GOODS-S (top) and GOODS-N (bottom) fields for the F770W (orange solid color), F1280W (orange circle hatch), and F1500W (orange line hatch) filters.  A small subset of the ancillary data available is shown for reference: Chandra X-ray 7 Ms imaging \citep[gray background region;][]{luo2017}, NIRCam F356W imaging as in DR 5 from multiple programs \citet[purple shaded region;][see]{Johnson2026}) with the JADES/NIRCam Deep and JADES/NIRCam Medium footprints (PID 1180) highlighted (purple solid line), JOF (PID 3215) \citep[blue violet solid line;][]{eisenstein2023}, SMILES MIRI 8-band survey \citep[red dashed line;][]{rieke2024, alberts2024a}, and HST WFC3 F160W footprint \citep[blue dotted line;][]{grogin2011}. }
    \label{fig:footprints}
\end{figure*}

\section{Survey Design and Acquisition}\label{sec:acq}

\subsection{JADES/MIRI Ultra-Deep Parallel}\label{sec:deep}

The JADES/MIRI ultra-deep coordinated parallel in GOODS-S (hereafter GS-Deep) is a $2\times2$ mosaic in a single filter, F770W, obtained in parallel with NIRCam ultra-deep imaging centered on the HUDF \citep[PID 1180;][]{rieke2023a, eisenstein2023a}.  Due to the arrangement of the instruments on the JWST focal plan and favorable position angles (PAs) for low near- and mid-IR backgrounds, the NIRCam observations placed the MIRI GS-Deep parallel in the southern part of the GOODS-S field ($\alpha, \delta=53.063469, -27.877115$), which is covered by a wealth of pre-JWST ancillary data (Figure~\ref{fig:footprints}, top).  The JADES survey was designed to extend its medium-depth NIRCam mosaic to this southern region, providing full coverage of the MIRI ultra-deep parallel with nearly two hours of exposure time in each of 8 NIRCam filters. To date, additional overlapping JWST data includes: deep NIRCam imaging obtained in parallel with deep NIRSpec MSA observations from PIDs 1210 \citep[]{bunker2024}, 3215 \citep[JADES Origins Field, JOF;][]{eisenstein2023} and 5997 (OASIS, PIs T. Looser \& F. D'Eugenio) as well as three-band NIRCam wide-field slitless spectroscopy (PID 4540, F. Sun et al., in prep). As a result, the MIRI GS-Deep parallel is covered by 9-15 bands of deep NIRCam imaging.  

Due to the scheduling of the JADES program, two MIRI pointings were observed for 61 ks and another two for 94 ks in Sept--Oct 2022; the remainder of the parallel observations were obtained in Sept--Oct 2023, bringing the total on-source exposure time to 155 ks per pointing, over a total area of 10.4 arcmin$^2$. The exposure time, depth (achieved and predicted by ETC), area, and coverage by NIRCam imaging programs is provided in Table~\ref{tbl:mosaic}. As NIRCam-MIRI parallels generate large data volumes, GS-Deep was obtained in the SLOWR1 readout mode with 57 groups, single integration, per dither (Table~\ref{tbl:exposure}).  In total, 114 exposures per pointing were obtained with a combination of the {\sc 9-POINT-WITH-MIRI-X} and {\sc 4-POINT-WITH-MIRI-X} subpixel dither patterns, which belong to a set of dither patterns optimized for NIRCam-MIRI parallels \citep{garcia-marin2016}.  {\sc X=F770W} was chosen in the first observing epoch.  The second epoch was switched to {\sc X=F1800W} to mitigate issues arising from small dither steps during background subtraction \citep{dicken2024, alberts2024a}. 

We note that our exposure time per integration ($\sim1300$ s; Table~\ref{tbl:exposure}) was chosen to maximize the ramp duration and obtain as much data as possible given the optimized NIRCam exposure setups and mechanism moves.  However, current guidelines\footnote{\url{https://jwst-docs.stsci.edu/mid-infrared-instrument/miri-observing-strategies/miri-cross-mode-recommended-strategies}} recommend maximum single integration exposure times of 300 s to minimize the number of pixels directly or indirectly effected by cosmic rays.  At $\sim1000$ s, it is expected that $\sim80\%$ of pixels have been affected$^1$; though partial ramps are recovered in the event of a cosmic ray hit, this may result in an increase in the uncertainty in ramp fitting and decrease the effective exposure time of the pixel.  However, any local or global affect appears largely mitigated in the GS-Deep MIRI parallel by the large number of dithers performed (see Section~\ref{sec:mosaics} for further discussion).

\subsection{JADES/MIRI Medium-Depth Parallels}\label{sec:medium}

The original setup and science goals of JADES/MIRI Medium parallels are briefly outlined in \citet{eisenstein2023a}.  We note here that the initially selected filter set (F770W, F1280W) was observed in GOODS-N but then modified in GOODS-S to split the time among three filters: F770W, F1280W, and F1500W.  

\subsubsection{GOODS-N}\label{sec:medium-gn}

JADES/MIRI medium depth parallels in GOODS-N (hereafter GN-Medium) were obtained in Feb 2023 over three contiguous pointings in the southern part of the field ($\alpha, \delta=189.18007, 62.139743$; Figure~\ref{fig:footprints}, bottom).  Exposures were obtained in two filters, F770W and F1280W, distributed over the available slots for $4-6$ NIRCam filter sets. All three pointings were observed in F770W for 6 ks.  F1280W was observed for 9 (6.6) ks for two (one) pointings. The area covered is 9.4 arcmin$^{2}$. The easternmost pointing (with 9 ks in F1280W) is covered by JADES NIRCam imaging; the two western pointings have $<30\%$ coverage from PID 4762 (PI S. Fujimoto).  GN-Medium overlaps fully with HST imaging. A summary of this information is provided in Table~\ref{tbl:mosaic}.

Again to mitigate large data volumes, JADES/MIRI GN-Medium was observed with the SLOWR1 readout mode.  The exposure setup can be found in Table~\ref{tbl:exposure}.  As in GS-Deep (Section~\ref{sec:deep}), the F770W exposures were obtained with longer ramps ($\sim500$ s) than currently recommended given the cosmic ray rate (see Section~\ref{sec:mosaics} for further discussion).
The dither pattern adopted was {\sc INTRAMODULEX} with three primary dithers and {\sc 2-POINT-WITH-MIRI-F1800W} subpixel dithers.

\subsubsection{GOODS-S}\label{sec:medium-gs}

Additional JADES/MIRI medium depth parallels (hereafter GS-Medium) were obtained in GOODS-S in Oct 2023 $-$ Jan 2024 in parallel with the JADES/NIRCam medium depth imaging of the northern HUDF/GOODS-S region. The original design included imaging in the F1280W and F1500W filters over three pointings that fell back on the JADES/MIRI GS-Deep parallel (Section~\ref{sec:deep}).  Due to the NIRCam field design, another three pointings would fall to the west of JADES/MIRI GS-Deep coverage, and so the time was split between F770W, F1280W, and F1500W\footnote{Due to the distribution of exposure times in NIRCam filter sets, we did not observe the MIRI filters in order of increasing wavelength, which is recommended to minimize persistence from the higher background at longer wavelengths.}. However, a series of guide star failures caused all but two of the JADES/NIRCam medium pointings with MIRI parallels to be rescheduled at different position angles than originally intended, which resulted in the configuration shown in Figure~\ref{fig:footprints} (top).  Efforts were made to place the MIRI parallels on existing or planned NIRCam data but full overlap was not possible without unacceptable modifications to the NIRCam survey design. A total of seven GS-Medium pointings\footnote{An eighth observation produced no viable data, see GS-Medium-2 description in Appendix~\ref{app:b}.} were observed with $11-39$ SLOWR1 groups and 6 dithers in combinations of F770W, F1280W, and F1500W, with exposures for the latter two broken into multiple integrations to avoid saturation (Table~\ref{tbl:exposure}). As discussed in the previous sections, the F770W exposures were obtained with longer ramps ($\sim900$ s) than currently recommended given the cosmic ray rate (see Section~\ref{sec:mosaics} for further discussion). A summary of the JADES/MIRI GS-Medium parallels is given in Table~\ref{tbl:mosaic}.   Additional details about each GS-Medium pointing are provided in Appendix~\ref{app:b}.

\begin{table*}[htb!]
    \footnotesize
    \centering
    \begin{tabular}{lcccccc}
    \hline
    \hline
        Filter & Exp Time & $5\sigma$ Depth & ETC v4 & Area & NIRCam & NIRCam F356W\\
         & [ks] & [$\mu$Jy] (AB) & [$\mu$Jy] (factor) & [arcmin$^2$] & PIDs & Overlap [$\%$] \\
    \hline
    \hline
    \multicolumn{7}{c}{\underline{\textbf{JADES/MIRI GS-Deep} ($\alpha, \delta=53.063469, -27.877115$)}}\vspace{1mm} \\
    F770W & 155 & 0.019 (28.2) & 0.05 & 10.4 & \makecell{1180, 1210, 1286, \\1287, 3215, 4540, \\5997} & $100\%$ \\
    \\
    \multicolumn{7}{c}{\underline{\textbf{JADES/MIRI GN-Medium} ($\alpha, \delta=189.18007, 62.139743$)}}\vspace{1mm} \\
    F770W  & 6.0 & 0.09 (26.5) & 0.2 & 9.5 & 1181, 4762 & $40\%$ \\
    F1280W & $6.6-9.0$ & $0.29-0.21$ ($25.2-25.6$) & 0.5-0.4 & 9.3 & $^{\prime\prime}$ & $^{\prime\prime}$ \\
    \\
    \multicolumn{7}{c}{\underline{\textbf{JADES/MIRI GS-Medium}}}\vspace{1mm} \\
    \multicolumn{7}{l}{\textbf{GS-Medium-1 ($\alpha, \delta=53.069574, -27.891897$):}}\vspace{1mm} \\
    F1280W & 13.5 & 0.22 (25.5) & 0.3 & 3.3 & 1180, 1210, 1286, 3215 &  100$\%$ \\
    F1500W & 10.9 & 0.39 (24.9) & 0.5 & $^{\prime\prime}$ & $^{\prime\prime}$ & $^{\prime\prime}$ \\
    \\
    \multicolumn{7}{l}{\textbf{GS-Medium-2 ($\alpha, \delta=52.995083, -27.861752$):}}\vspace{1mm} \\
    F770W  & 5.6 & 0.10 (26.3) & 0.2 & 3.3 &  1286, 3215 & $100\%$  \\
    F1280W  & 10.9 & 0.27 (25.3) & 0.4 & $^{\prime\prime}$ & $^{\prime\prime}$ & $^{\prime\prime}$  \\
    F1500W  & 13.5 & 0.37 (25.0) & 0.4 & $^{\prime\prime}$ & $^{\prime\prime}$ & $^{\prime\prime}$ \\
    \\
    \multicolumn{7}{l}{\textbf{GS-Medium-3 ($\alpha, \delta=53.019449, -27.797543)$:}}\vspace{1mm} \\
    F770W  & 5.0 & 0.10 (26.4) & 0.2 & 3.3 & 1287, 2514, 3990 & $100\%$ \\
    F1280W  & 10.9 & 0.24 (25.5) & 0.4 & $^{\prime\prime}$ & $^{\prime\prime}$ & $^{\prime\prime}$ \\
    F1500W  & 10.9 & 0.36 (25.0) & 0.5 & $^{\prime\prime}$ & $^{\prime\prime}$ & $^{\prime\prime}$ \\
    \\
    \multicolumn{7}{l}{\textbf{GS-Medium-4} ($\alpha, \delta$=52.963641, -27.701178):}\vspace{1mm} \\
    F770W  & 5.0 & 0.10 (26.4) & 0.2 & 3.3 & $-$ & $-$  \\
    F1280W  & 10.9 & 0.24 (25.4) & 0.4 & $^{\prime\prime}$ & $-$ & $-$  \\
    F1500W  & 10.9 & 0.34 (25.1) & 0.5 & $^{\prime\prime}$ & $-$ & $-$ \\
    \\
    \multicolumn{7}{l}{\textbf{GS-Medium-5} ($\alpha, \delta$=53.069689, -27.701076):}\vspace{1mm} \\
    F770W  & 5.0 & 0.11 (26.3) & 0.2 & 3.3 & 1180, 1286 & $92\%$ \\
    F1280W  & 10.9 & 0.23 (25.5) & 0.4 & $^{\prime\prime}$ & $^{\prime\prime}$ & $^{\prime\prime}$ \\
    F1500W  & 5.0-10.9 & $0.51-0.33$ ($24.6-25.1$) & $0.7-0.5$ & 5.6 & $^{\prime\prime}$ & $95\%$ \\
    \\
    \multicolumn{7}{l}{\textbf{GS-Medium-6 ($\alpha, \delta=53.107764, -27.670805$):}}\vspace{1mm} \\
    F1500W  & 5.0 & 0.5 (24.7) & 0.7 & 3.2 & $-$ & $-$ \\
    \hline
    \end{tabular}
    \caption{Summary of JADES/MIRI parallel observations (Section~\ref{sec:acq}). Notes: Column 3: The depth is given as the aperture-corrected, $5\sigma$ point source sensitivity. Column 4: Predicted $5\sigma$ point source sensitivities from ETC v4.  Column 6-7: Summary of overlap between the MIRI parallels and NIRCam imaging programs.  PIDs are listed for overlapping NIRCam programs that obtained $\ge6$ filters and have $>10\%$ overlap with the MIRI mosaic.  The percentage covered is determined based on the F356W imaging that is included in JADES DR5. The list of NIRCam programs providing overlap to the MIRI parallels is not exhaustive.  NIRCam program references: 1180, 1181, 1210, 1286, 1287 \citep[JADES;][]{eisenstein2023a, Johnson2026, Robertson2026}; 3215, 4540 \citep[JOF;][]{eisenstein2023}; 2514 \citep[PANORAMIC;][]{williams2025}; 3990 \citep[BEACON;][]{morishita2025}, 5997 (OASIS, PIs T. Looser \& F. D'Eugenio)  }
    \label{tbl:mosaic}
\end{table*}

\section{Data Processing}\label{sec:data_reduction}

The data processing for the JADES/MIRI parallels was performed following the data reduction pipeline presented for the SMILES MIRI imaging mosaic in \citet{alberts2024a}.  Here we provide an overview and note any differences from what is described in that work.

\subsection{Overview}\label{sec:proc_overview}

Reduction of the JADES MIRI parallels was done using {\sc v1.16.1} of the {\sc JWST Calibration Pipeline} \citep{bushouse2023} with Calibration Reference Data System (CRDS)\footnote{\url{https://jwst-crds.stsci.edu/}} {\sc v12.0.8} and CRDS context 1303.  The {\sc Pipeline} is divided into three stages to which we added custom routines for addressing artifacts from persistence, removing warm pixels, custom background subtraction, and astrometry corrections.  We started from the uncalibrated ({\sc uncal}) files obtained from the Mikulski Archive for Space Telescopes (MAST)\footnote{\url{https://mast.stsci.edu/portal/Mashup/Clients/Mast/Portal.html}}.

\begin{figure*}[tbh!]
    \centering
    \includegraphics[width=1.6\columnwidth]{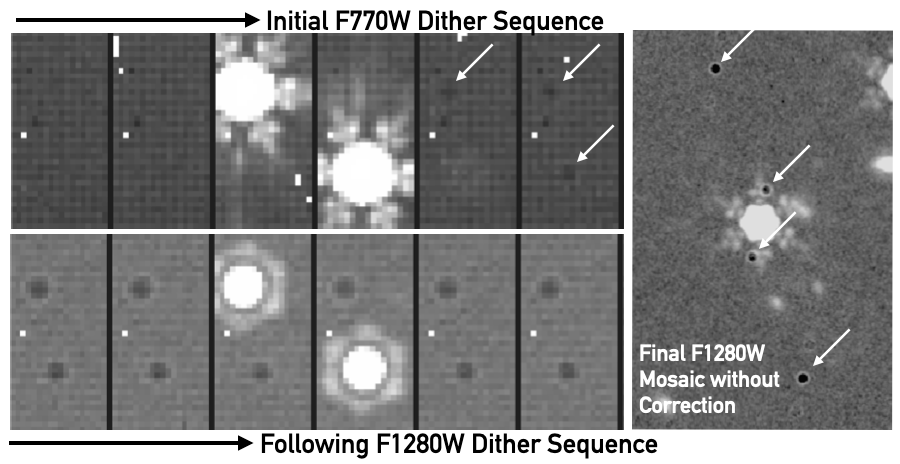}
   
    \caption{Artifacts created from persistence due to  saturated pixels (Section~\ref{sec:persistence}).  (left) A small area of the MIRI imaging detector is shown for six dithers for F770W (top), which was followed by six dithers in F1280W (bottom).  An under-exposed artifact appears following the observation where a bright source saturates multiple pixels in at least one group. (Right) Without any correction, the final F1280W mosaic shows these persistence artifacts prominently. }

    \label{fig:persistence}
\end{figure*}

Prior to running the {\sc Pipeline}, we manually flagged persistence due to a small number of saturated pixels; our procedure is described in Section~\ref{sec:persistence}.  Stage 1 ({\sc calwebb\_detector1}) of the {\sc Pipeline} was then run to perform processing at the detector-level on raw ramp data, applying corrections for non-ideal detector behaviors as described in \citet{morrison2023}.  This stage includes flagging cosmic rays and cosmic rays showers, which we expect to affect a large portion of the pixels in GS-Deep due to our long exposure times (Sections~\ref{sec:deep},~\ref{sec:mosaics}).

Processed slope images (aka {\sc rate} and {\sc rateint} files) were then fed to Stage 2 ({\sc calwebb\_image2}) which associates a WCS solution with individual exposures and then applies flat fielding and flux calibrations\footnote{This step would also perform background subtraction, which we skipped in favor of our custom routine (see Section~\ref{sec:bkg_sub}).}.  Stage 2 {\sc cal} files were then median stacked in detector space to identify and mask warm/hot pixels \citep[see][]{alberts2024a} which were not flagged in the bad pixel mask applied in Stage 1.  The number of warm/hot pixels on the MIRI detectors is known to be increasing with time\footnote{\url{https://jwst-docs.stsci.edu/known-issues-with-jwst-data/miri-known-issues/miri-imaging-known-issues}}, but at a slow enough rate that they can be identified by stacking roughly contemporary exposures. 
To test this, we looked at the distribution of warm pixels over a $1-3$ month baseline using the GS-Medium data and found that the changes are negligible.  Therefore, we grouped the GN-Medium and GS-Medium exposures to generate their respective warm/hot pixel masks.  For GS-Deep, we separately generated two warm pixel masks for the 2022 and 2023 data.

The Stage 2 {\sc cal} files were then background subtracted as described in Section~\ref{sec:bkg_sub} and astrometry corrections were applied according to available ancillary data (Section~\ref{sec:astrometry}). Final mosaicking was done though the {\sc Pipeline} Stage 3 step ({\sc calwebb\_image3}), which performs final outlier detection and resampling.  In this stage, we skipped the built-in astrometry correction in lieu of custom corrections (with the exception of GS-Medium-4 and -6, see Section~\ref{sec:astrometry}) and additionally skipped the skymatch step as it is redundant with our background subtraction technique (Section~\ref{sec:bkg_sub}).  The final mosaics are described in Section~\ref{sec:mosaics}.

\subsubsection{Persistence Artifacts}\label{sec:persistence}

Persistence, residual charge in pixels exposed to bright or saturating sources, is common in infrared detectors.  For MIRI, persistence can be caused not only by bright or saturating astrophysical sources, sky backgrounds, or cosmic rays incident on the imager plane during a given observing program, but also by such sources being seen by the detector prior to that program, during previous observations with any instrument or during slews\footnote{This is due to the fact that the MIRI detectors are continuously exposed to the sky and clocking.} \citep[see][for details]{dicken2024}.  This makes identifying persistence in a systematic way very challenging.  To add insult to injury, the magnitude and decay timescale of persistence are strongly dependent on several factors that make it highly difficult to model \citep{morrison2023, dicken2024}.  As such, the only avenue for dealing with persistence artifacts currently is manually identifying and masking the pixels affected.

A small number of persistence artifacts were present in our initial reductions.  An example is shown in Figure~\ref{fig:persistence}; in the final mosaic (far right), this artifact manifested as a distinctive light donut with a dark core, repeated according to the dither pattern.  In analyzing this artifact, it was determined that the cause is a bright star with a small number of saturated pixels in its core.  The first 6 dithers in F770W for GN-Medium are shown in Figure~\ref{fig:persistence} (left), where we display a part of the detector where the star first appears in the third dither, takes a small step in the fourth dither and then moves back out of the frame.  Initially, a weak positive persistence is visible.  This decays rapidly \citep{dicken2024}, but is followed by a negative persistence, where the pixels display reduced sensitivity and are under-luminous.  This second stage of persistence has an unknown timescale; in this example, the artifact is seen throughout the observation, including through the subsequent F1280W exposures, where the bright star is no longer saturating. The total exposure time is $\sim3.5$ hours; however, we note that our setup repeated dither positions such that this source hit the same pixels twice in F770W and then twice in F1280W. 

\begin{figure}[tbh!]
    \centering
    
    \includegraphics[width=0.95\columnwidth]{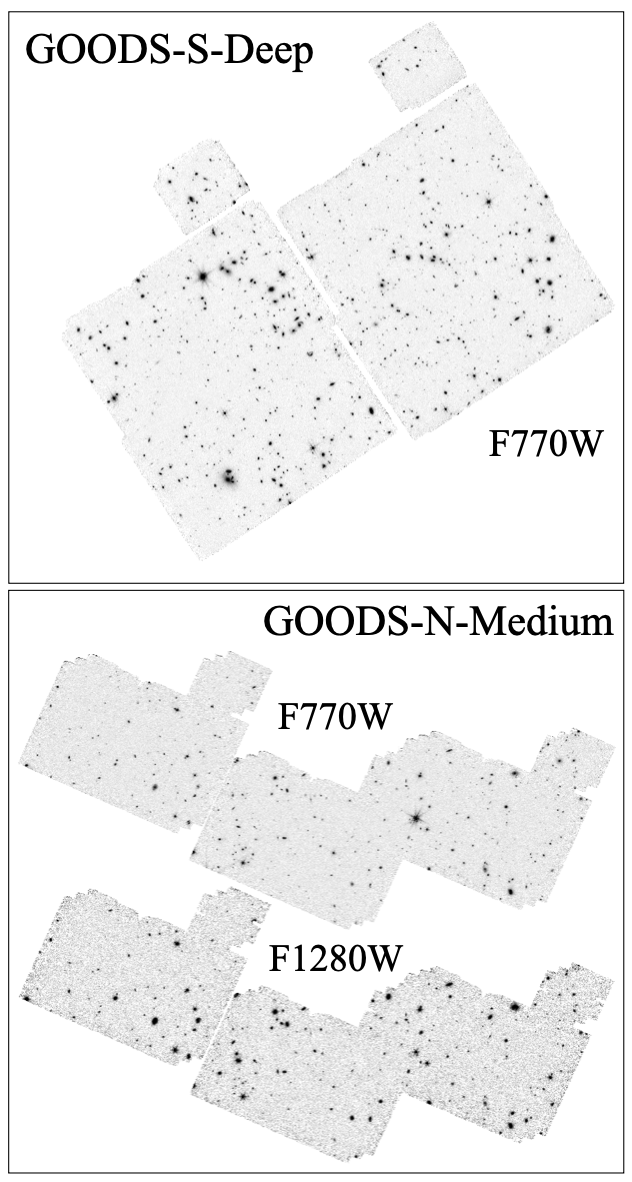}
    \caption{Greyscale images of GS-Deep F770W (top) and the GN-Medium F770W and F1280W (bottom) mosaics.  Due to the significantly different exposure times (155 vs $6-9$ ks), these are not shown on the same scale.  GN-Medium and GS-Medium (Figures~\ref{fig:mosaics2} and ~\ref{fig:mosaics3}) are displayed on the same scale and stretch, derived from the GN-Medium F770W.  }

    \label{fig:mosaics1}
\end{figure}

To remove these artifacts associated with saturation from the final mosaic, we did the following: prior to running Stage 1 (Section~\ref{sec:proc_overview}), we ran a time-ordered set of {\sc uncal} files through the first two steps of {\sc calwebb\_detector1} to initiate the Data Quality array\footnote{\url{https://jwst-docs.stsci.edu/accessing-jwst-data/jwst-science-data-overview}} and run the saturation flagging. Saturated pixels\footnote{Defined as having less than three unsaturated groups in the pixel ramp.} were then tracked and masked in all \textit{subsequent} exposures in the set.  In other words, if a pixel experiences saturation in the first dither, it was set to {\sc DO\_NOT\_USE} in the {\sc PixelDQ} in all subsequent dithers. We did not mask the pixel in the exposure in which it saturates because 1) the persistence signal is small compared to the source, 2) it is already flagged as saturated, and 3) it will be recorded as a {\sc NaN} or a partial ramp will be recovered, if possible.  We propagated the flagged pixels from the first filter (i.e. F770W) to any additional filters (i.e. F1280W and/or F1500W) regardless of whether the pixels were saturated in the longer wavelength bands\footnote{Because our exposure setup repeated dither positions, prior to flagging pixels for persistence we checked if a given a pixel is $>3$ the median of all pixels to avoid flagging source pixels.}.  For GN-Medium, the percentage of pixels flagged was $\sim0.05-0.13\%$ as our exposure setup was largely successful in avoiding saturation.  Similar percentages are masked in the GOODS-S pointings.

\begin{figure*}[tbh!]
    \centering
    
    \includegraphics[width=1.5\columnwidth]{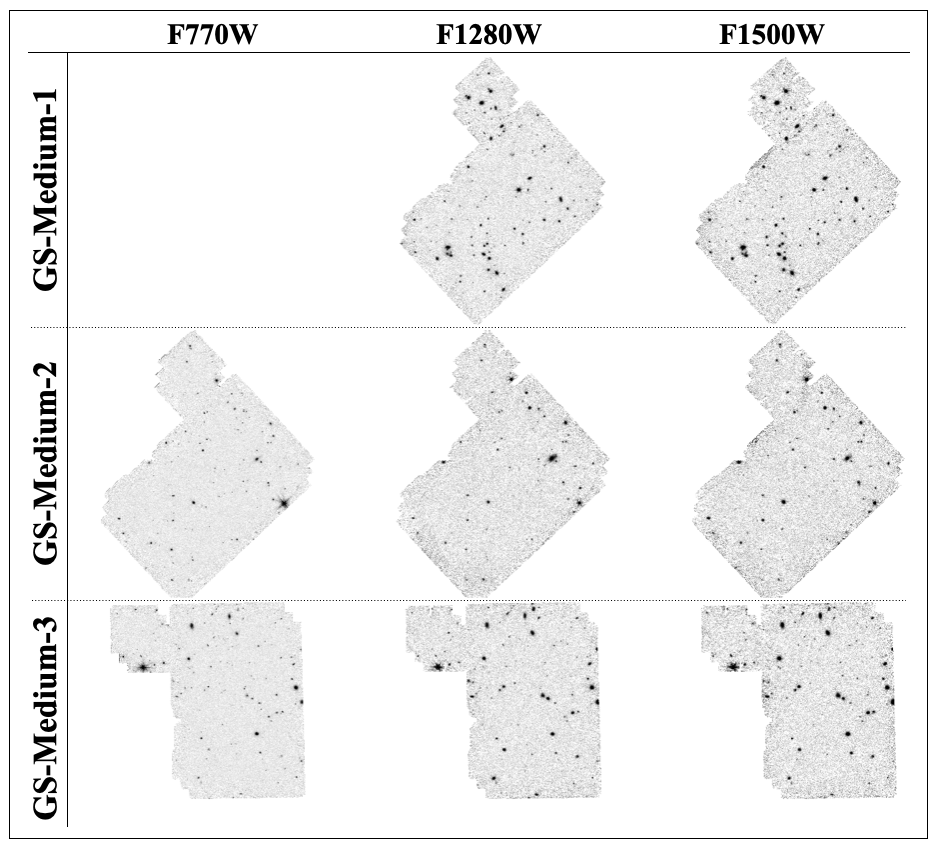}
    \caption{Greyscale images of GS-Medium-1, -2, and -3 in F770W (left), F1280W (middle), and F1500W (right).  Scale and strength match GN-Medium F770W (Figure~\ref{fig:mosaics1}).}
    \label{fig:mosaics2}
\end{figure*}

We note again that this procedure does not address persistence from sources prior to our observations.  We have one obvious example in GS-Medium-3 F1280W and F1500W: persistence obtained during slewing across a bright source manifests as six parallel streaks in the final mosaic (reflecting the dither pattern).  This persistence faded in 6 hours and is not evident in the F770W imaging, which was observed last.  At this time, we have not implemented manual masking for such cases.

\subsubsection{Background Subtraction}\label{sec:bkg_sub}

\begin{figure*}[tbh!]
    \centering
    
    \includegraphics[width=1.5\columnwidth]{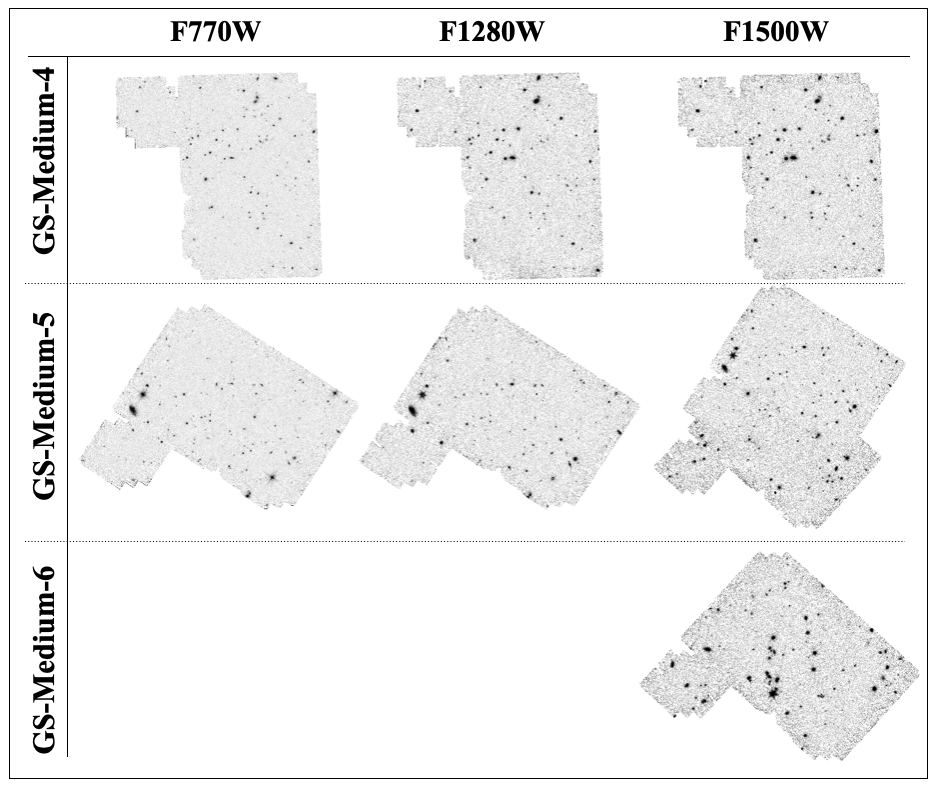}
    \caption{Greyscale images of GS-Medium-4, -5, and -6 in F770W (left), F1280W (middle), and F1500W (right).  Scale and strength match GN-Medium F770W (Figure~\ref{fig:mosaics1}).}
    \label{fig:mosaics3}
\end{figure*}

Our background subtraction procedure is fully described in \citet[][]{alberts2024a}.  In summary, we first constructed robust source masks through an iterative process aimed at capturing both faint sources and faint extended flux. For each contemporary dataset spanning multiple, independent pointings, a super background was then constructed by masking the sources and image edges and subtracting the global background level in each {\sc cal} image and then taking a pixel-wise median of a stack of all masked and modified {\sc cal} files in detector space. This super background was then scaled back to the global background level and subtracted from each {\sc cal} file.  The global background level was determined via a simple median of all pixels after masking sources and the image edges and we note that it is known to vary even from dither to dither. Finally, (masked) median filtering was performed along rows/columns and any remaining large 2D background gradients were removed. 

As demonstrated in \citet[][]{alberts2024a}, super background subtraction results in deeper mosaics than predicted by the ETC (Table~\ref{tbl:mosaic}).  Combining contemporaneous exposures over multiple, spatially separated pointings provides for robust estimates of residual detector plus sky background per pixel and mitigates `shadowing' caused by oversubtraction from dither steps being too small to `step over' extended sources \citep{dicken2024}.  This effect was especially apparent in the 2022 GS-Deep F770W imaging, which used the {\sc 9-POINT-WITH-MIRI-F770W} subpixel dithers.  Subsequent JADES/MIRI parallels adopted larger dither patterns.

The mid-IR sky background, however, does have a temporal component, as found for the SMILES mosaic \citep{alberts2024a}.  For our datasets with observations taken over long timescales, we grouped contemporaneous observations on timescales of a few months and iterated to find the balance between minimizing the effect of the changing background against the benefits of combining more spatially distinct pointings. For GS-Deep, super backgrounds were constructed separately for the Oct 2022 and Oct 2023 observations. GN-Medium was all grouped as it was all obtained together. For GS-Medium, we divided the F1500W exposures into those observed in Oct/Nov and Jan 2023.  However, for F770W and F1280W, only one pointing each was obtained in Jan, so we grouped these with the Oct/Nov observations to avoid the shadowing effect. According to the JWST Backgrounds Tool\footnote{\url{https://jwst-docs.stsci.edu/jwst-other-tools/jwst-backgrounds-tool}}, the F770W pointing obtained in Jan was observed with a $\sim9\%$ higher background at 7.7$\,\mu$m as compared to those obtained in Oct/Nov. The result is further discussed in Section~\ref{sec:mosaics}.

\subsubsection{Astrometry Correction}\label{sec:astrometry}

Astrometry corrections were applied pointing-by-pointing as described in \citet{alberts2024a}.  For GS-Deep and GN/GS-Medium pointings with sufficient overlapping NIRCam (Table~\ref{tbl:mosaic}), the astrometry is registered to JADES \citep[which is registered to GAIA, see][]{Johnson2026} with a typical accuracy of $0.01\arcsec$ ($1\sigma$), a tenth of the MIRI native pixel size.  For the two GN-Medium pointings without NIRCam overlap, we corrected the astrometry using the CHARGE HST catalog (G. Brammer priv. comm.) with an accuracy of $0.015\arcsec$ ($1\sigma$), reflecting the fewer matches between HST and MIRI sources. The GS-Medium pointings with NIRCam coverage were corrected to $0.01-0.015\arcsec$ ($1\sigma$). For GS-Medium-4 and -6, which lack NIRCam overlap, we ran the {\sc TweakReg} step in Stage 3 of the {\sc Pipeline} with the standard GAIA catalog; however, we cannot verify the quality of these corrections due to the low number of quality source matches between GAIA and MIRI and the lack of a well registered JWST or HST catalog.

\subsection{Final Mosaics}\label{sec:mosaics}

Figure~\ref{fig:rgb} shows a F770W, F444W, F200W RGB image of GS-Deep and Figures~\ref{fig:mosaics1}--\ref{fig:mosaics3} show grayscale images the GS-Deep, GN-Medium and GS-Medium fields. Exposure time maps are shown in Appendix~\ref{app:a} in Figure~\ref{fig:exp_time} for GS-Deep and GN-Medium and Figure~\ref{fig:exp_time2} for GS-Medium.  As can be seen in the Figure~\ref{fig:exp_time}, overlap between the pointings and the Lyot coronagraph in GS-Deep provides a small area with total exposure times of up to $\sim300$ ks.  Our final mosaics have a pixel scale of $0.05998939\arcsec$, twice the pixel size of the NIRCam GOODS-S mosaics\footnote{The JADES NIRCam pixel size in GOODS-S is $0.02999476\arcsec$, approximately the native pixel size for the short-wavelength channel and closely matched to the empirical measurements of the pixel size of archival HST mosaics \citep{Johnson2026}}, and are oriented such that North is up and East is to the left. The coordinate system is ICRS. Overlapping pointings in different filters are registered to the same pixel grid and image dimensions for ease of use.  We note that GS-Medium-1 specifically is re-projected onto the grid and larger image dimensions of GS-Deep.  Image pixels with no data are set to NaN. 

We note that, as found in previous MIRI surveys \citep{yang2023a, alberts2024a}, the {\tt Pipeline}-generated error images underestimate the uncertainties by factors of $2-3$.  To assess the depth of our mosaics, we measure the median rms of boxes placed across the mosaics themselves using the {\sc photutils} {\sc Background2D} routine after masking sources and the mosaic edges.  Aperture-corrected $5\sigma$ point source sensitivities are reported in Table~\ref{tbl:mosaic} for circular apertures with $r=0.42, 0.42$ and $0.49\arcsec$ for F770W, F1280W, and F1500W, respectively.  These apertures correspond to the $65\%$ encircled energy (EE) derived from (semi-)empirical MIRI point spread functions (PSFs)\footnote{The $65\%$ EE at F770W and F1280W are similar due to broadening of the MIRI PSF from the cruciform artifact at F770W \citep{gaspar2021}.}, described in \citet{alberts2024a}.  An additional factor of $1.5\times$ is applied to account for correlated pixel noise from resampling to smaller pixels (L. Constanin, priv. comm.).  The point source sensitivities reached by JADES/MIRI are 19 nJy (28.2 AB, $5\sigma$) at F770W for GS-Deep and range from 90 nJy (26.5 AB) to 510 nJy (24.6 AB) for GN- and GS-Medium.  As in \citet{alberts2024b}, we find that our mosaic depths are $\sim1.3-2.5$ $\times$ deeper than the ETC predictions. Completeness simulations and number counts GS-Deep, spanning roughly five orders of magnitude at F770W through the combination of JADES and SMILES, were presented in \citet{stone2024a}.  JADES GS-Deep F770W was found to be $\sim70\%$ complete at its $5\sigma$ detection limit (Table~\ref{tbl:mosaic}).

\begin{figure}[tbh!]
    \centering
    
    \includegraphics[width=0.95\columnwidth]{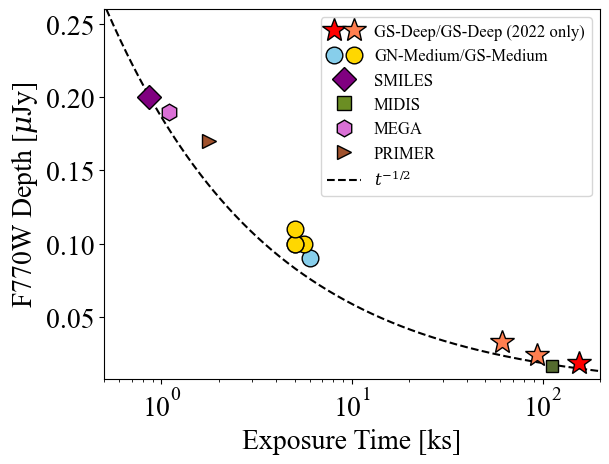}
    \caption{The F770W $5\sigma$ point source sensitivities reached as a function of exposure time in ks for various MIRI surveys.  GS-Deep is shown as both the final mosaic (red star) and the partial data collected in 2022 (two pointings with 61 ks and two pointings with 94 ks, shown as orange stars).  The 2023 data reached the same depths.  The JADES GN-Medium, JADES GS-Medium, SMILES \citep{alberts2024a}, MIDIS \citep{rinaldi2025}, MEGA \citep{backhaus2025}, and PRIMER \citep{liu2026a} F770W depths are shown as a blue circle, yellow circles, a purple diamond, a green square, pink hexagon, and brown triangle, respectively.  The dashed line shows the the expectation that sensitivity scales with the square root of time (normalized to SMILES).   }

    \label{fig:rms_vs_exptime}
\end{figure}

Finally, we examine whether there is evidence that our mosaics were compromised by the large fraction of pixels presumably impacted by cosmic rays.  As discussed in Section~\ref{sec:proc_overview}, $80\%$ of pixels are expected to be directly or indirectly affected by a cosmic ray in a $\sim1000$ s single-integration exposure.  The GS-Deep and GS-Medium F770W integrations range from $\sim800-1300$ s, while the GN-Medium F770W pointings were obtained with integrations time of $\sim500$ s, still in excess of the recommendation (Table~\ref{tbl:exposure}). In Figure~\ref{fig:rms_vs_exptime}, we show the F770W image depths ($5\sigma$) as a function of total exposure time for JADES/MIRI and SMILES.  Normalizing to SMILES F770W (216 s integrations), we find that these depths reached scale as the square root of the exposure time between SMILES and JADES GS-Deep, as expected.  We further break GS-Deep into just the original 2022 data, which was obtained in two sets of two pointings with 61 and 94 ks of exposure time (Section~\ref{sec:deep}). The 2022-only GS-Deep, GN- and GS-Medium mosaics fall slightly above this trend set by SMILES and the full GS-Deep, which could be related to the high cosmic ray rate, which is then mitigated by the high redundancy (114 total exposures per pointing) of the full GS-Deep dataset.  However, an alternative explanation is that these mosaics have slightly less optimal background subtraction due to having fewer independent pointings to combine into the super background and/or grouping non-contemporary pointings (Section~\ref{sec:bkg_sub}) as was necessary for some GS-Medium pointings due to guide star failures (Section~\ref{sec:medium-gs}).  For comparison, we also show the depths reported by the MIDIS \citep{rinaldi2025}, MEGA \citep[PID 3794;][]{backhaus2025}, and PRIMER \citep[PID 1837;][]{liu2026a} surveys, which were obtained with shorter ramps ($\lesssim400$ s) and are generally in good agreement with the expected trend. This comparison is approximate given that different data reduction pipelines were used\footnote{We also not that the F770W depths reported by MIDIS, MEGA, and PRIMER were measured in slightly smaller apertures ($r=0.25-0.35\arcsec$) than used for JADES and SMILES ($r=0.42\arcsec$)}.

\section{Data Release}\label{sec:data_release}

All fully reduced JADES/MIRI mosaics in GOODS-N and GOODS-S are included in JADES Data Release 5 and are available as High Level Science Products (HLSPs) on MAST at \url{https://archive.stsci.edu/hlsp/jades}.  In total, this includes 36.4, 25.8, and 22 arcmin$^2$ in F770W, F1280W, and F1500W, respectively, ranging from medium ($0.09-0.51\,\mu$Jy [26.5-24.6 AB]) to ultra-deep ($0.019\,\mu$Jy [28.2 AB]) $5\sigma$ point source sensitivities.  The DR5 MIRI mosaics have a pixel size of $0.05998939\arcsec$ and overlapping MIRI filters are registered to the same grid.  The fits files include five image extensions: 

\begin{itemize}
    \item {\sc SCI}: science extension in units of MJy sr$^{-1}$
    \item {\sc ERR}: resampled per pixel uncertainty estimates, as output by the {\sc Calibration Pipeline}, also in MJy sr$^{-1}$
    \item {\sc WHT}: relative weight map of the pixels, as output by the {\sc Calibration Pipeline}
    \item {\sc EXT}: an exposure time map in seconds showing the sum of the exposure time contributing to each pixel, not accounting for ramps shortened by jumps
    \item {\sc NIM}: the number of individual exposures that contribute to each pixel
\end{itemize} 

NIRCam prior-based forced photometry in all MIRI filters is included in the full JADES DR 5 catalog, providing measurements corresponding to all JADES/NIRCam sources.  In areas without NIRCam coverage, MIRI photometry is provided based on a MIRI detection image.  For full details, see \citet{Robertson2026}.

\vspace{5mm}
\textit{Data Availability:} The data presented in this article was obtained from the Mikulski Archive for Space Telescopes (MAST) at the Space Telescope Science Institute. The JADES data can be accessed at \dataset[doi: 10.17909/8tdj-8n28]{https://archive.stsci.edu/doi/resolve/resolve.html?doi=10.17909/8tdj-8n28}.  The JADES HLSP Data Releases can be accessed at \url{https://archive.stsci.edu/hlsp/jades}.

\begin{acknowledgments}

The JADES team thanks the MIRI instrument team for their work in developing, commissioning, and supporting MIRI and the JWST pipeline team for their development of the JWST Calibration Pipeline. The authors thank Andr\'as G\'aspar for his work on the empirical MIRI PSFs and Mihai Cara for assistance in writing astrometry correction software.  SA, JL, and GHR acknowledge support from the JWST Mid-Infrared Instrument (MIRI) Science Team Lead, grant 80NSSC18K0555, from NASA Goddard Space Flight Center to the University of Arizona.  DJE, KH, JMH, ZJ, BDJ, MR, BR, and CNAW acknowledge support from the NIRCam Science Team contract to the University of Arizona, NAS5-02105. AJB acknowledges funding from the ``First Galaxies'' Advanced Grant from the European Research Council (ERC) under the European Union's Horizon 2020 research and innovation program (Grant agreement No. 789056). ECL acknowledges support of an STFC Webb Fellowship (ST/W001438/1). PGP-G acknowledges support from grant PID2022-139567NB-I00 funded by Spanish Ministerio de Ciencia e Innovaci\'on MCIN/AEI/10.13039/501100011033, FEDER, UE. ST acknowledges support by the Royal Society Research Grant G125142. The research of CCW is supported by NOIRLab, which is managed by the Association of Universities for Research in Astronomy (AURA) under a cooperative agreement with the National Science Foundation. This work is based on observations made with the NASA/ESA/CSA James Webb Space Telescope. The data were obtained from the Mikulski Archive for Space Telescopes at the Space Telescope Science Institute, which is operated by the Association of Universities for Research in Astronomy, Inc., under NASA contract NAS 5-03127 for JWST.  The authors acknowledge use of the {\it lux} supercomputer at UC Santa Cruz, funded by NSF MRI grant AST 1828315.
\end{acknowledgments}

%

\vspace{5mm}
\facilities{JWST, HST}


\software{JWST Calibration Pipeline, astropy, photutils, sep}



\clearpage

\appendix
\renewcommand\thefigure{\thesection.\arabic{figure}} 
\renewcommand\thetable{\thesection\arabic{table}}

\section{Exposure Setups and Exposure Time Maps for the JADES/MIRI Mosaics}\label{app:a}
\setcounter{figure}{0}
\setcounter{table}{0}

\begin{figure*}[tbh!]
    \centering
    \includegraphics[width=1.8\columnwidth]{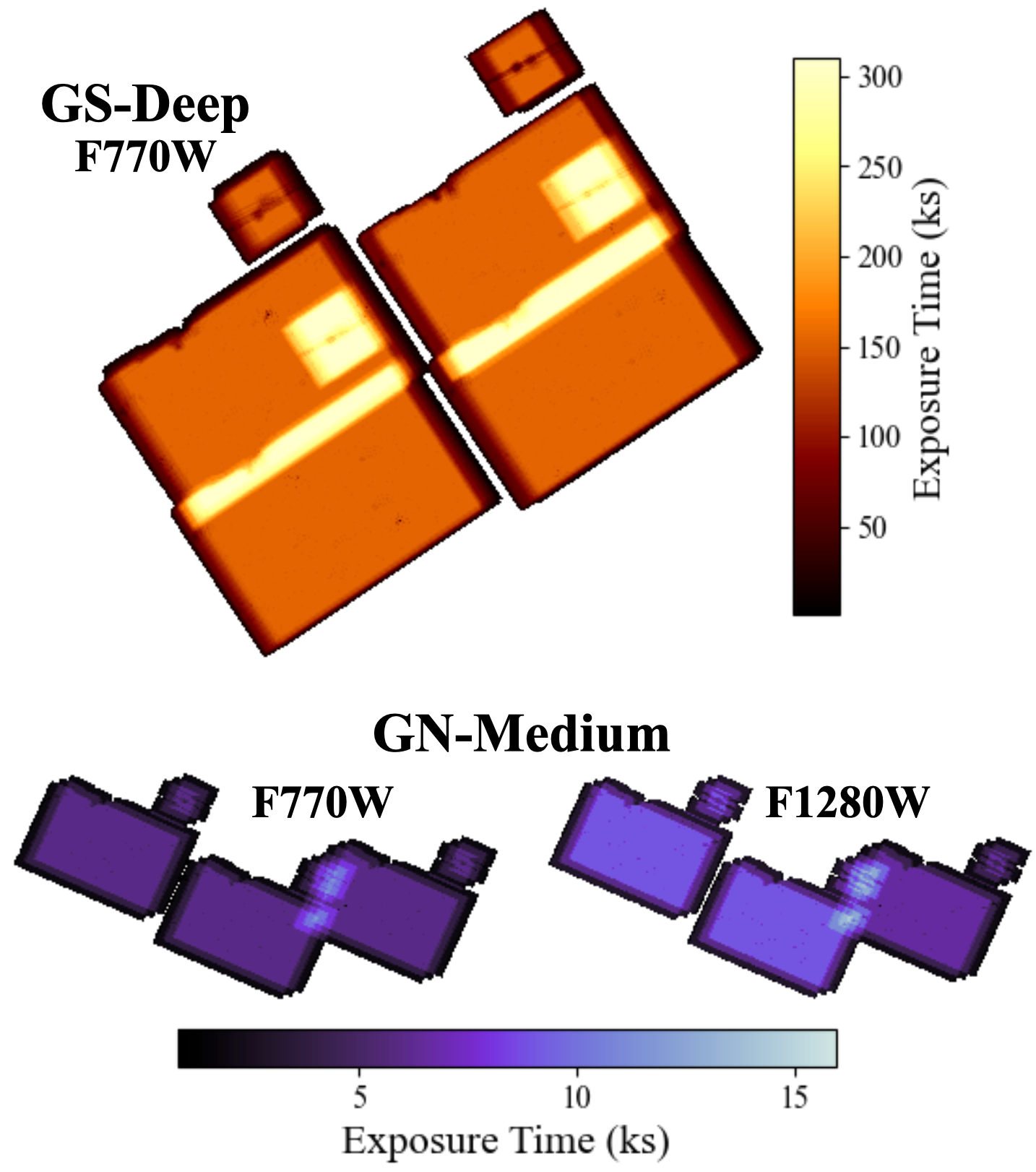}
    \caption{The exposure time maps for the GS-Deep (top) and GN-Medium (bottom) pointings. }

    \label{fig:exp_time}
\end{figure*}

\begin{figure*}[tbh!]
    \centering
    \includegraphics[width=1.8\columnwidth]{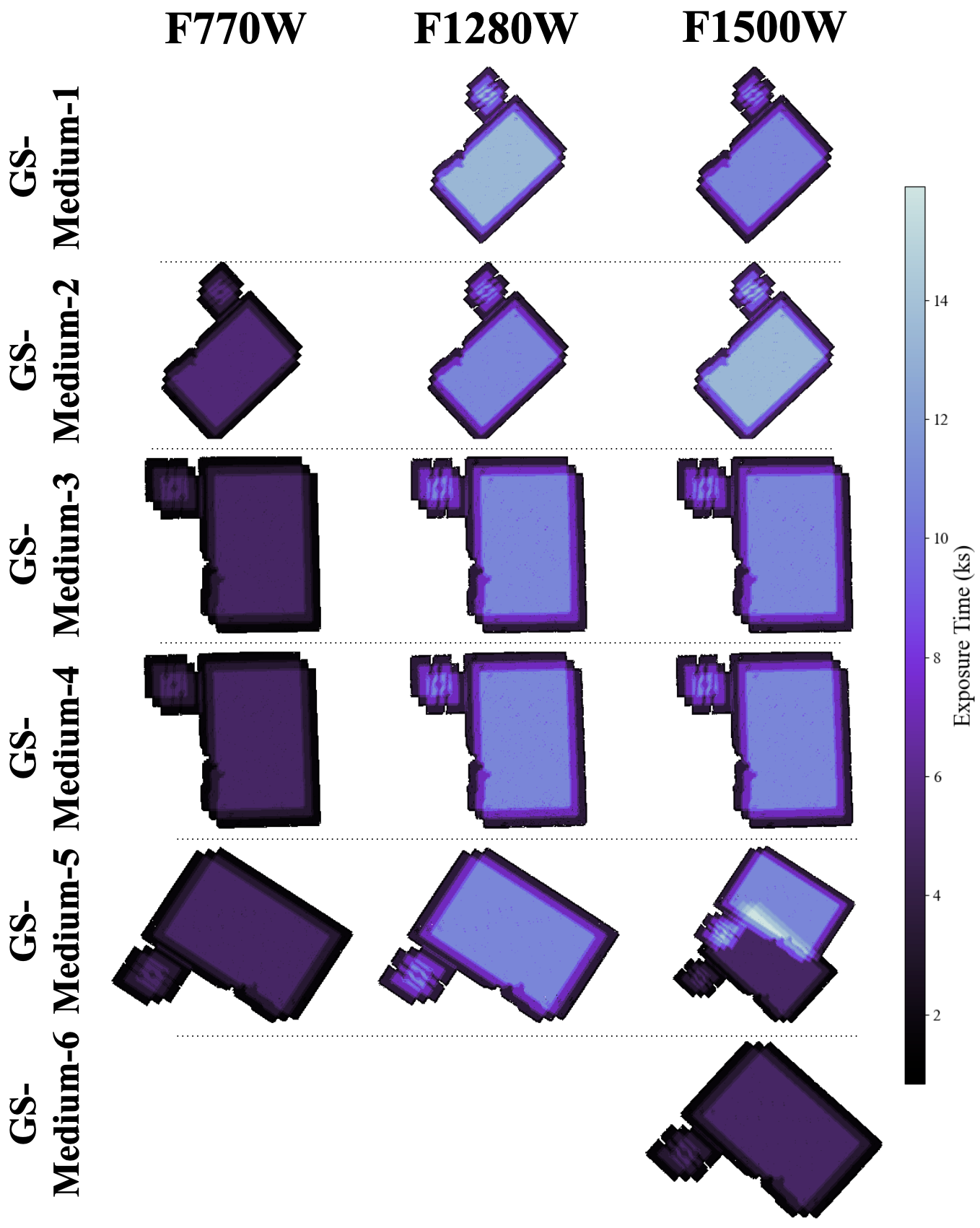}
    \caption{The exposure time maps for the GS-Medium pointings.}

    \label{fig:exp_time2}
\end{figure*}

In this appendix, we provide the exposure setup and exposure time maps for the JADES/MIRI GS-Deep, GN-Medium, and GS-Medium parallels, described in Section~\ref{sec:acq} and Table~\ref{tbl:mosaic}.  In Table~\ref{tbl:exposure}, we list the number of pointings,  exposure parameters (groups, integrations), number of dithers per NIRCam filter set, and number of total exposures per pointing.  The exposure time maps in ks are shown for GS-Deep and GN-Medium in Figure~\ref{fig:exp_time} and for the GS-Medium mosaics in Figure~\ref{fig:exp_time2}.  

\begin{table*}[h!]
    \centering
    \footnotesize
    \begin{tabular}{lcccccc}
    \hline
    \hline
        Filter & \# of  & $N_{\rm groups}$ & $t_{\rm int}$ & $N_{\rm int}$ & Dithers per & Exposures per  \\
         & Pointings &  & [s] & & Filter Set & Pointing  \\
    \hline
    \hline
    \multicolumn{6}{c}{\underline{MIRI GS-Deep}} \\
    F770W$^a$ & 4 & 57 & 1362 & 1 & 9 & 90 \\
    F770W$^a$ & 4 & 57 & 1362 &  1 & 4 & 12 \\
    \\
    \multicolumn{6}{c}{\underline{MIRI GN-Medium}} \\
    F770W & 3 & 21 & 502 & 1 & 6 & 12 \\
    F1280W & 3 & 10 & 234 & 2 & 6 & 12-18 \\
    \\
    \multicolumn{6}{c}{\underline{MIRI GS-Medium}} \\
    \multicolumn{6}{l}{\textbf{GS-Medium-1:}}\\
    F1280W & 1 & 15 & 358 & 3 & 6 & 12 \\
    F1500W & 1 & 12 & 287 & 3 & 6 & 12 \\
    \\
    \multicolumn{6}{l}{\textbf{GS-Medium-2:}}\\
    F770W & 1 & 39 & 932 & 1 & 6 & 6 \\
    F1280W & 1 & 12 & 287 & 3 & 6 & 12 \\
    F1500W & 1 & 15 & 358 & 3 & 6 & 12 \\
    \\
    \multicolumn{6}{l}{\textbf{GS-Medium-3:}}\\
    F770W & 1 & 39 & 932 & 1 & 6 & 6 \\
    F1280W & 1 & 12 & 287 & 3 & 6 & 12 \\
    F1500W & 1 & 12 & 287 & 3 & 6 & 12 \\
    \\
    \multicolumn{6}{l}{\textbf{GS-Medium-4:}}\\
    F770W & 1 & 39 & 932 & 1 & 6 & 6 \\
    F1280W & 1 & 12 & 287 & 3 & 6 & 12 \\
    F1500W & 1 & 12 & 287 & 3 & 6 & 12 \\
    \\
    \multicolumn{6}{l}{\textbf{GS-Medium-5:}}\\
    F770W & 1 & 35 & 836 & 1 & 6 & 6 \\
    F1280W & 1 & 12 & 287 & 3 & 6 & 12 \\
    F1500W$^d$ & 1 & 12 & 287 & 3 & 6 & 12 \\
    F1500W$^d$ & 1 & 11 & 263 &  3 & 6 & 6 \\
    \\
    \multicolumn{6}{l}{\textbf{GS-Medium-6:}}\\
    F1500W & 1 & 11 & 263 & 3 & 6 & 6 \\
    \\
    \hline
    \end{tabular}
    \caption{The exposure setups for GS-Deep, GN-Medium, and GS-Medium. Total exposure times, depth, area, and overlap with existing NIRCam programs for each mosaic is presented in Table~\ref{tbl:mosaic}.  Notes-- Column 3: the number of SLOWR1 groups per integration. Column 4-5: exposure time in a single integration and number of integration. Column 6: Dithers per filter set is the number of MIRI dithers per NIRCam filter set. Column 7: Exposures per pointing is the total number of MIRI exposures over all NIRCam filter sets per pointing. $^a$Fully overlapping, two dither patterns were used to match the prime NIRCam observations. $^b$Partially overlapping due to a guide star failure (Figure~\ref{fig:footprints}, Section~\ref{sec:medium-gs}).}
    \label{tbl:exposure}
\end{table*}

\renewcommand\thefigure{\thesection.\arabic{figure}} 
\renewcommand\thetable{\thesection\arabic{table}}

\section{Additional details on the JADES/MIRI GOODS-S Medium Parallels}\label{app:b}
\setcounter{figure}{0}
\setcounter{table}{0}

\vspace{1mm}
\underline{\textbf{GS-Medium-1 ($\alpha, \delta=53.069574, -27.891897$):}} A single pointing in F1280W and F1500W was $\sim80\%$ completed in Oct 2023, suffering a guide star failure before finishing the F1500W exposure.  This is the only pointing to successfully overlap the JADES/MIRI GS-Deep F770W parallel as originally intended and is fully covered by NIRCam imaging from multiple programs. 

\vspace{1mm}
\underline{\textbf{GS-Medium-2 ($\alpha, \delta=52.995083, -27.861752$):}} A single pointing just west of the GS-Deep and GS-Medium-1 successfully observed in Oct 2023 in F770W, F1280W, and F1500W. 
This pointing has $\sim100\%$ overlap with NIRCam imaging from multiple programs, with JADES (PID 1180, 1268) and JOF (PID 3215) providing multi-band coverage. GS-Medium-2 was preceded by a failed MIRI observation, where imaging in F1500W was acquired (PID 1180, obs 23) when MIRI continued to expose for $\sim1.9$ hr after a guide star failure terminated the prime observations.  We have verified that this data was not dithered.  A cursory analysis suggested the pointing was not stable during each exposure and so we opted to exclude these observations from our reduction of GS-Medium-2.

\vspace{1mm}
\underline{\textbf{GS-Medium-3 ($\alpha, \delta=53.019449, -27.797543)$:}} A single pointing located $\sim4$ arcmin north of GS-Medium-2, and was obtained in parallel with a replanned NIRCam observation in Nov 2023.  F770W, F1280W, and F1500W were observed successfully and have $\sim100\%$ overlap with NIRCam imaging from multiple programs, including JADES (PID 1287) and pure parallel programs PANORAMIC \citep[PID 2514;][]{williams2025}, and BEACON \citep[PID 3990;][]{morishita2025}.

\vspace{1mm}
\underline{\textbf{GS-Medium-4} ($\alpha, \delta$=52.963641, -27.701178):}. A single pointing located $\sim6.5$ arcmin to the northwest of GS-Medium-3, observed in parallel with a replanned NIRCam pointing  in Nov 2024. Full observations were obtained in F770W, F1280W, and F1500W.  There is currently no CANDELS HST or NIRCam coverage of this pointing.

\vspace{1mm}
\underline{\textbf{GS-Medium-5 ($\alpha, \delta=53.069689, -27.701076$):}} Two partially overlapping pointings located $\sim10.5$ arcmin north of GS-Deep, observed in parallel with two re-planned NIRCam pointings in Jan 2024.  One successfully observed F770W, F1280W, and F1500W.  A second partially overlapping pointing obtained a short 5 ks exposure in F1500W prior to a guide star failure.  GS-Medium-5 is almost entirely ($\sim90-95\%$) overlapped by JADES NIRCam imaging (PID 1180 and 1286) and HST.

\vspace{1mm}
\underline{\textbf{GS-Medium-6 ($\alpha, \delta=53.107764, -27.670805$):}} A short exposure obtained in F1500W in parallel with a replanned NIRCam observation in Jan 2024, cut short by a guide star failure. This pointing is just north of GS-Medium-5, but is outside current NIRCam coverage, though it is partially covered by HST.

\clearpage


\bibliography{main}{}
\bibliographystyle{aasjournal}



\end{document}